\documentclass[aps,prd,reprint,amsmath,amssymb,preprintnumbers,superscriptaddress,nofootinbib,longbibliography]{revtex4-1}
\usepackage[colorlinks]{hyperref}
\usepackage{graphicx}

\newcommand{\J}{\text{J}}
\newcommand{\MP}{M_\text{P}}

\begin{document}

\preprint{CTPU-18-44}
\preprint{LDU2018-06}
\preprint{OU-HET-1002}
\preprint{UMN--TH--3913/19}
\preprint{FTPI--MINN--19/04}

\title{
\large
Hillclimbing inflation in metric and Palatini formulations
}
\renewcommand{\thefootnote}{\alph{footnote}}

\author{
Ryusuke Jinno
}
\affiliation{
Center for Theoretical Physics of the Universe, Institute for Basic Science (IBS), Daejeon 34126, Korea
}

\author{
Kunio Kaneta
}
\affiliation{
William I.~Fine Theoretical Physics Institute, School of Physics and Astronomy, University of Minnesota, Minneapolis, MN 55455, USA
}

\author{
Kin-ya Oda
}
\affiliation{
Department of Physics, Osaka University, Osaka 560-0043, Japan
}

\author{
Seong Chan Park 
}
\affiliation{
Department of Physics \& IPAP, Yonsei University, Seoul 03722, Korea
}

\begin{abstract}
A new setup of cosmic inflation with a periodic inflaton potential and conformal factor is discussed 
in the metric and Palatini formulations of gravity.
As a concrete example, we focus on a natural-inflation-like inflaton potential, 
and show that the inflationary predictions fall into the allowed region of cosmic microwave background observations in both formulations.
\end{abstract}

\maketitle

\section{Introduction}
\label{sec:Introduction}

Cosmic inflation~\cite{Starobinsky:1980te,Guth:1980zm,Sato:1980yn} is one of the 
most important pillars in modern cosmology.
It not only solves the flatness and horizon problems~\cite{Guth:1980zm} 
and dilutes possible unwanted relics~\cite{Sato:1980yn},
but also successfully produces the primordial density perturbations necessary for late-time cosmic structures~\cite{
Mukhanov:1981xt,Kodama:1985bj}.
Cosmic microwave background (CMB) observations have been contributing to narrow down inflationary models
(e.g. Ref.~\cite{Akrami:2018odb}),
and further improvement in the experimental sensitivity is expected
in the future~\cite{Matsumura:2013aja,Inoue:2016jbg,Delabrouille:2017rct}.

As the observational constraints become more stringent,
there have been growing interests in the attractors of inflation models.
One of the well-known attractors is inflation with the $R^2$ term~\cite{Starobinsky:1980te},
which realizes inflation through a modification of the 
gravitational action.
Another well known example is the inflation with nonminimal coupling,
which has been studied from 80's~\cite{Lucchin:1985ip,Futamase:1987ua,Salopek:1988qh,Fakir:1990eg,Makino:1991sg,vanderBij:1993hx,vanderBij:1994bv, Park:2008hz},
and been more extensively studied after Higgs inflation~\cite{Salopek:1988qh,CervantesCota:1995tz}
became one of the main topics in modern particle cosmology~\cite{Bezrukov:2007ep, Hamada:2014iga, Hamada:2014wna, He:2018mgb}.\footnote{
The setup can be generalized to a requirement 
that $V_\J/\Omega^2$ approaches a constant value at the range of inflation~\cite{Park:2008hz}.
}
These attractor-type models enjoy their inflationary predictions well within the observational sweet spot.
Such attractor models have been generalized to the $\alpha$-attractor~\cite{Kallosh:2013yoa}.
In this context, a systematic study of small conformal-factor attractor has been proposed in~Ref.~\cite{Jinno:2017jxc},
and as a result a new realization of Higgs inflation has been discovered~\cite{Jinno:2017lun}.

In this paper we introduce a new attractor setup in the context of hillclimbing inflation. 
The model consists of periodic conformal factor (i.e. scalar field coupling to the Ricci scalar) and potential,
but it still enjoys the attractor-type inflationary predictions.
One of the interesting features of the model is that it predicts the observationally favored values of inflationary observables
both in the metric and the Palatini formulations of gravity~\cite{Palatini1919,Einstein1925}.\footnote{
The paper~\cite{Palatini1919} by Palatini is often referred to for the Palatini formulation,
but it is reported~\cite{Ferraris1982} that the formulation was introduced in the paper~\cite{Einstein1925} by Einstein.
}
The Palatini formulation, where the connection $\Gamma$ is regarded as another independent variable 
to take variation with in addition to the metric $g$, has been known as a formulation of gravity
distinctive from the metric formulation, in which $\Gamma(g)$ is set to be the Levi-Civita connection and is given as a function of $g$.
Though these two formulations give the identical predictions
as long as we consider the pure Einstein gravity and minimally-coupled matter fields,
it is known that they lead to different dynamics once gravity action is modified or 
nontrivial couplings between gravity and matter are introduced.
This difference has been attracting considerable attention recently~\cite{Bauer:2008zj,Bauer:2010jg,
Rasanen:2017ivk,Tenkanen:2017jih,Racioppi:2017spw, Racioppi:2018zoy, Antoniadis:2018ywb,Carrilho:2018ffi,Enckell:2018hmo,Enckell:2018kkc,Jarv:2017azx,Markkanen:2017tun,Rasanen:2018fom,Rasanen:2018ihz,Takahashi:2018brt},
especially in the context of Higgs inflation.

The organization of the paper is as follows.
First we introduce our setup in Sec.~\ref{sec:Setup} where the inflationary predictions are analytically investigated both in the metric and the Palatini formulations.
Next we present numerical results in Sec.~\ref{sec:Results}.
We finally conclude in Sec.~\ref{sec:Conclusion}.

\section{Setup}
\label{sec:Setup}

The setup we consider in this paper is
\begin{align}
S
&= 
\int d^4x \sqrt{-g_\J}
\left[
\frac{1}{2} \Omega R_\J
- \frac{\kappa}{2} g_\J^{\mu \nu} \partial_\mu \phi_\J \partial_\nu \phi_\J - V_\J(\phi_\J)
\right],
\label{eq:Setup}
\end{align}
with the ``natural inflation''-type potential\footnote{
Note that the potential (\ref{eq:VJ}) is realized for example as the difference of two $\cos$-type potentials:
\begin{align}
\sin^4(\phi_\J/f) 
&= 
\frac{1 - \cos(2\phi_\J/f)}{2}
- \frac{1 - \cos(4\phi_\J/f)}{8}.
\end{align}
This type of potential has been used for example in Ref.~\cite{Czerny:2014xja},
and its UV completion has been discussed for example in Ref.~\cite{Czerny:2014wza}.
As will become clear in the following discussion, $V_\J$ needs to behave as $\phi_\J^4$ around a potential minimum in order for inflation to occur in the Palatini formulation.
}
\begin{align}
V_\J
&=
\Lambda^4
\sin^4
\left( \frac{\phi_\J}{f} \right).
\label{eq:VJ}
\end{align}
We adopt the Planck unit $\MP \equiv 1/\sqrt{8 \pi G} = 1$ (with $G$ being the Newtonian constant) throughout the paper.
Also, the subscript ``J" refers to the Jordan frame.
The main purpose of the present paper is to point out that this type of potential predicts the observationally 
allowed region with the following choice of the conformal factor:\footnote{
{
The form of $\Omega$ is chosen from phenomenological requirement $V_\J/\Omega^2 \to \text{const.}$ 
at a point of $V_\J=0$ as discussed in Ref.~\cite{Jinno:2017jxc}.
The existence of coincident zero of $\Omega$ with the $V_\J=0$ point is motivated by the generalized multiple-point principle: 
this model fits into the class of ``new'' version of MPP corresponding to the weak energy condition~\cite{odakinTalk}.
}
}
\begin{align}
\Omega
&= 
\omega
\sin^2 
\left( \frac{\phi_\J}{\eta f} \right).
\label{eq:Omega}
\end{align}
The normalization constant $\omega=1/\sin^2(\pi / \eta)$ is 
fixed to give a proper value of gravitational interaction $\Omega=1$ at the minimum of the potential $\phi_\J = \pi f$,
and $\eta$ is a real free parameter controlling the ratio between the period of 
the potential~($2\pi f$) and that of the conformal factor~($2\pi \eta f$). 
We further impose $\eta > 1$ in order for inflation to successfully end at the potential minimum.
Also, because of the subtleties discussed in Appendix~\ref{app:Limit}, 
we further restrict $\eta$ to be $\eta \geq 2$. 
The results for $1 < \eta < 2$ are explained in this appendix.

\begin{figure}[t]
\begin{center}
\includegraphics[width=0.9\columnwidth]{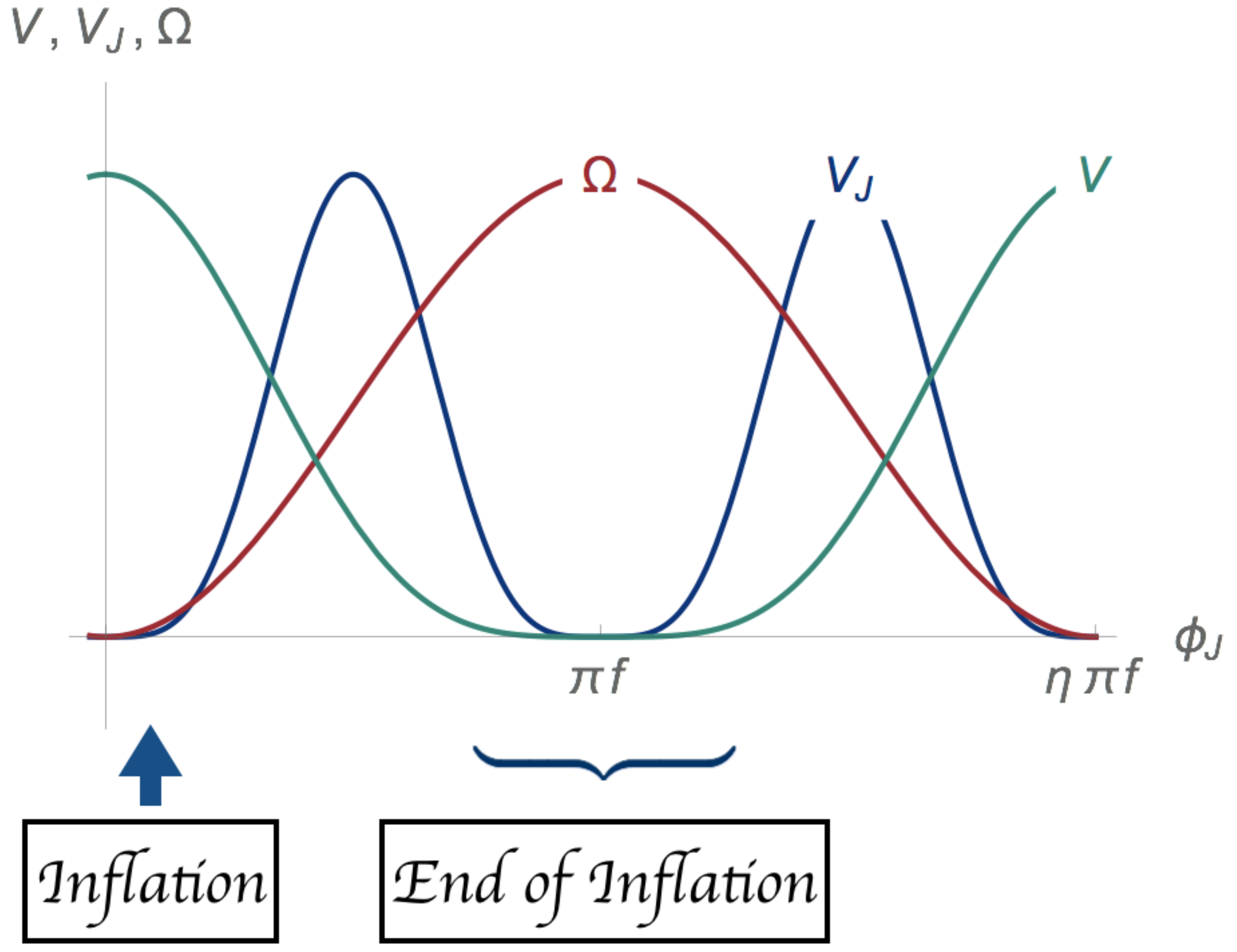}
\caption{\small
Illustration of the setup.
Conformal factor $\Omega$ (red),
Jordan-frame potential $V_\J$ (blue), 
and Einstein-frame potential $V = V_\J/\Omega^2$ (green).
The vertical scale is arbitrary.
In this plot we took $\eta = 2$.
}
\label{fig:SetupJordan}
\end{center}
\end{figure}

The setup is summarized in Fig.~\ref{fig:SetupJordan}.
As we will see later, one of the interesting features of the present setup is that 
it predicts the observationally allowed region both in the metric and Palatini formulations of gravity.
Also, for the metric formulation, 
the existence of the kinetic term in the action (\ref{eq:Setup}) is not even mandatory to achieve successful inflation.
Therefore we consider both $\kappa = 0$ and $\kappa=1$ for the metric case,
while we exclusively take $\kappa = 1$ for the Palatini case.

In the following discussion, we redefine the metric:
\begin{align}
g_{\mu \nu}
&= 
\Omega g_{\J  \mu \nu}.
\label{eq:metric}
\end{align}
This gives the transformation law of the Ricci scalar, which depends on the choice of the formulation.
It also gives the relation between the original inflaton $\phi_\J$ and the canonical inflaton $\phi$ after the redefinition.
The resulting Einstein-frame action becomes
\begin{align}
S
&=
\int d^4x \sqrt{-g}
\left[
\frac{1}{2} R
- \frac{1}{2} (\partial \phi)^2 - V(\phi)
\right],
\end{align}
with $(\partial \phi)^2 \equiv g^{\mu \nu} \partial_\mu \phi \partial_\nu \phi$.
The Einstein-frame potential $V$ in terms of the original inflaton $\phi_\J$ becomes
\begin{align}
V
&=
\frac{V_\J}{\Omega^2}
=
\frac{\Lambda^4 \sin^4(\phi_\J/  f)}{\omega^2 \sin^4(\phi_\J/ \eta f)}.
\end{align}
This is one of the hillclimbing setups~\cite{Jinno:2017jxc}, namely,
around $\phi_\J = 0$, 
the potential behaves like $V_\J \propto \phi_\J^4$
while the conformal factor behaves like $\Omega \propto \phi_\J^2$.
As a result, as we see from Fig.~\ref{fig:SetupJordan}, 
the potential minimum around $\phi_\J = 0$  is no more a minimum in the Einstein frame.
At the same time, the relation between the Jordan and Einstein frame inflatons gives
an exponential stretching in the canonical field in the latter frame.
Here the relation between the original and new inflatons $\phi_\J$ and $\phi$ depends 
on the formulation of gravity and the choice of the kinetic term $\kappa$.
As a result, our basic scenario is as follows:
\begin{itemize}
\item
First, inflation occurs around $\phi_\J = +0$ (corresponding to $\phi = \infty$), which is no more a minimum
but a maximum in the Einstein frame.
\item
Next, the inflaton rolls down to the minimum $\phi_\J = \pi f$ (corresponding to $\phi = 0$),
where the inflation ends.
\end{itemize}
The slow-roll parameters and $e$-folding are calculated in terms of  the original inflaton as
\begin{align}
\epsilon_V
&\equiv
\frac{1}{2} 
\left(
\frac{dV/d\phi}{V}
\right)^2
= 
\frac{1}{2} 
\left(
\frac{dV/d\phi_\J}{V}
\right)^2
\frac{1}{(d\phi/d\phi_\J)^2},
\\
\eta_V
&\equiv 
\frac{d^2V/d\phi^2}{V}
\nonumber \\
&= 
\left[
\frac{d^2V/d\phi_\J^2}{V}
\frac{1}{(d\phi/d\phi_\J)^2}
-
\frac{dV/d\phi_\J}{V}
\frac{d^2\phi/d\phi_\J^2}{(d\phi/d\phi_\J)^3}
\right],
\\
N
&= 
\int
\frac{d\phi}{\sqrt{2\epsilon_V}}
=
\int 
\frac{d\phi_\J}{d\ln V/d\phi_\J}
\left(
\frac{d\phi}{d\phi_\J}
\right)^2.
\end{align}
We take the lower end of the integration to be $\phi=0$ so that $N=0$ at $\phi=0$.\footnote{
Note that the difference between this definition and the usual definition $N = 0$ at $\max(|\epsilon_V|, |\eta_V|) = 1$
gives only next-leading corrections in $N$ to the inflationary predictions.
}

The inflationary predictions become
\begin{align}
{\mathcal P}_\zeta
&\simeq
\frac{1}{24\pi^2}\frac{V}{\epsilon_V},
~~~~
n_s
\simeq
1 - 6\epsilon_V + 2\eta_V,
~~~~
r 
\simeq
16\epsilon_V.
\end{align}
In the following analysis we fix ${\mathcal P}_\zeta = 2.1 \times 10^{-9}$~\cite{Akrami:2018odb}.
This gives the overall height of the potential $\propto\Lambda^4$ as a function of $f$ for each setup.
In the following we examine the metric and the Palatini formulations one by one.

\subsection*{Metric formulation}

The redefinition of the metric (\ref{eq:metric}) gives the relation between the 
Ricci scalars in the Jordan and the Einstein frames:
\begin{align}
R_\J
&=
\Omega
\left[
R
+ 3 \Box \ln \Omega
- \frac{3}{2} (\partial \ln \Omega)^2
\right].
\label{eq:RJ}
\end{align}
A new contribution to the kinetic term appears from the last term in this expression.
As a result, the canonical inflaton $\phi$ is related with the field in the Jordan frame
\begin{align}
\frac{d\phi}{d\phi_\J}
&=
- \sqrt{
\frac{\kappa}{\Omega^2}
+ 
\frac{3}{2} \left(
\frac{d\ln \Omega}{d\phi_\J}
\right)^2
},
\label{eq:dphidphiJMetric}
\end{align}
with the boundary condition $\phi = 0$ at $\phi_\J = \pi f$.
Here we put the minus sign so that the field value during inflation $\phi_\J = +0$ corresponds to $\phi = \infty$ (see Fig.~\ref{fig:SetupJordan}).

\subsubsection*{Case $\kappa = 0$}

\begin{figure}[t]
\begin{center}
\includegraphics[width=\columnwidth]{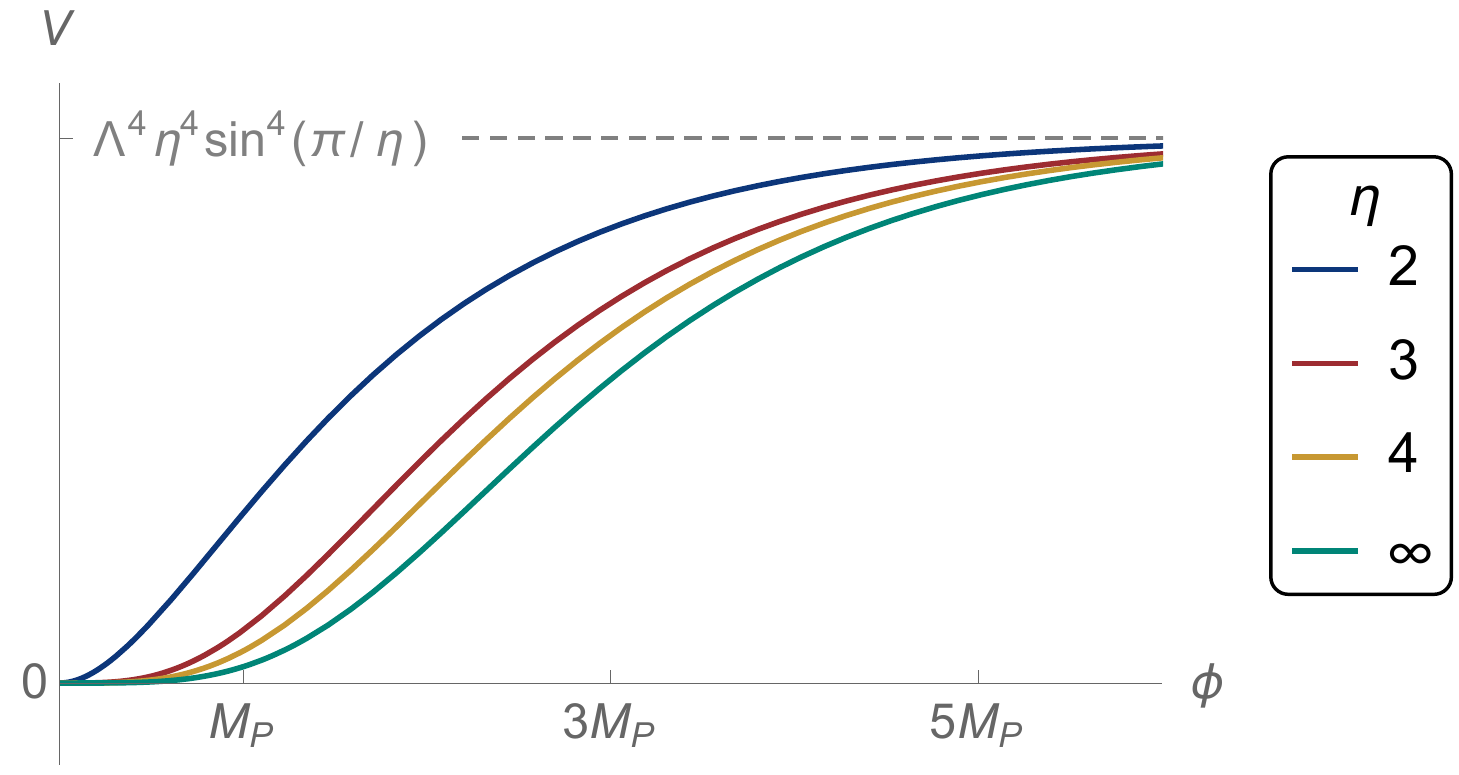}
\caption{\small
Einstein-frame potential as a function of the canonical inflaton $\phi$
in the metric formulation without the kinetic term ($\kappa = 0$).
The blue, red, yellow and green (from top to bottom) lines are for $\eta = 2$, $3$, $4$ and $\infty$, respectively.
Note that $\phi = 0$ corresponds to $\phi_\J = \pi f$, while $\phi = \infty$ corresponds to $\phi_\J = + 0$
(see Fig.~\ref{fig:SetupJordan} and Eq.~(\ref{eq:dphidphiJMetric}) or (\ref{eq:dphidphiJPalatini})).
}
\label{fig:SetupEinstein1}
\end{center}
\end{figure}

We first consider the case without the inflaton kinetic term in the Jordan frame ($\kappa = 0$).\footnote{
{
This setup can be mapped to an $F(R_\J)$ setup by solving the constraint equation for $\phi_\J$ in the Jordan frame:
$\frac{1}{2}R_\J d\Omega/d\phi_\J= dV_\J/d\phi_\J$ or $R_\J = 2\left(dV_\J/d\phi_\J\right)/\left(d\Omega/d\phi_\J\right) \equiv {\cal G}(\phi_\J)$. By feeding $\phi_\J ={\cal G}^{-1}(R_\J)$ back into the original action, we get an equivalent $F(R_\J)$ theory
with $F(R_\J) = \frac{1}{2}\Omega({\cal G}^{-1}(R_\J))R_\J - V_\J({\cal G}^{-1}(R_\J))$.
}
}
The inflaton $\phi$ is still dynamical due to the contribution found in Eq.~(\ref{eq:RJ}).
In this case, we can solve Eq.~(\ref{eq:dphidphiJMetric}) analytically:
\begin{align}
\phi
&=
- \sqrt{\frac{3}{2}} \ln \Omega
~~
\leftrightarrow 
~~
\Omega
=
e^{- \sqrt{\frac{2}{3}}\phi}.
\end{align}
Using Eq.~(\ref{eq:Omega}), we can replace $\phi_\J$ in terms of $\phi$ 
in the expression for the potential:
\begin{align}
V
&= 
\Lambda^4 
e^{2\sqrt{\frac{2}{3}} \phi}
\sin^4
\left[
\eta
\arcsin
\left[
\sin
\left(
\frac{\pi}{\eta}
\right)
e^{- \frac{1}{2} \sqrt{\frac{2}{3}} \phi}
\right]
\right].
\label{eq:VMetric_kappa=0}
\end{align}
We see that the inflationary predictions depend only on the periodicity ratio $\eta$, not on $f$.
The potential (\ref{eq:VMetric_kappa=0}) is plotted in Fig.~\ref{fig:SetupEinstein1}.

The blue, red, yellow, and green lines correspond to $\eta = 2$, $3$, $4$, and $\infty$, respectively.
To understand the potential shape, it is instructive to expand the potential assuming $e^{- \sqrt{2/3} \phi} \ll 1$:
\begin{align}
V
&\simeq
\Lambda^4 
\eta^4 
\sin^4 
\left(
\frac{\pi}{\eta}
\right)
\nonumber \\
&~~~~
\times
\left[
1 - \frac{2}{3}(\eta^2 - 1) \sin^2 \left( \frac{\pi}{\eta} \right) e^{- \sqrt{\frac{2}{3}} \phi} + \cdots
\right].
\label{eq:VapproxMetric_kappa=0}
\end{align}
From this expression we see that the potential indeed develops an exponentially flat plateau for $\phi \gg 1$.
Also, for $\eta \to \infty$, the potential (\ref{eq:VMetric_kappa=0}) becomes simpler
\begin{align}
V
&\to
\Lambda^4 
e^{2\sqrt{\frac{2}{3}} \phi}
\sin^4
\left[
\pi
e^{- \frac{1}{2} \sqrt{\frac{2}{3}} \phi}
\right]
~~
{\rm for}
~~
\eta \to \infty.
\label{eq:VetainfMetric_kappa=0}
\end{align}
Later we see that this potential coincides with the one in the Palatini formulation with a specific choice of $f$.

\begin{figure*}[t]
\begin{center}
\includegraphics[width=\columnwidth]{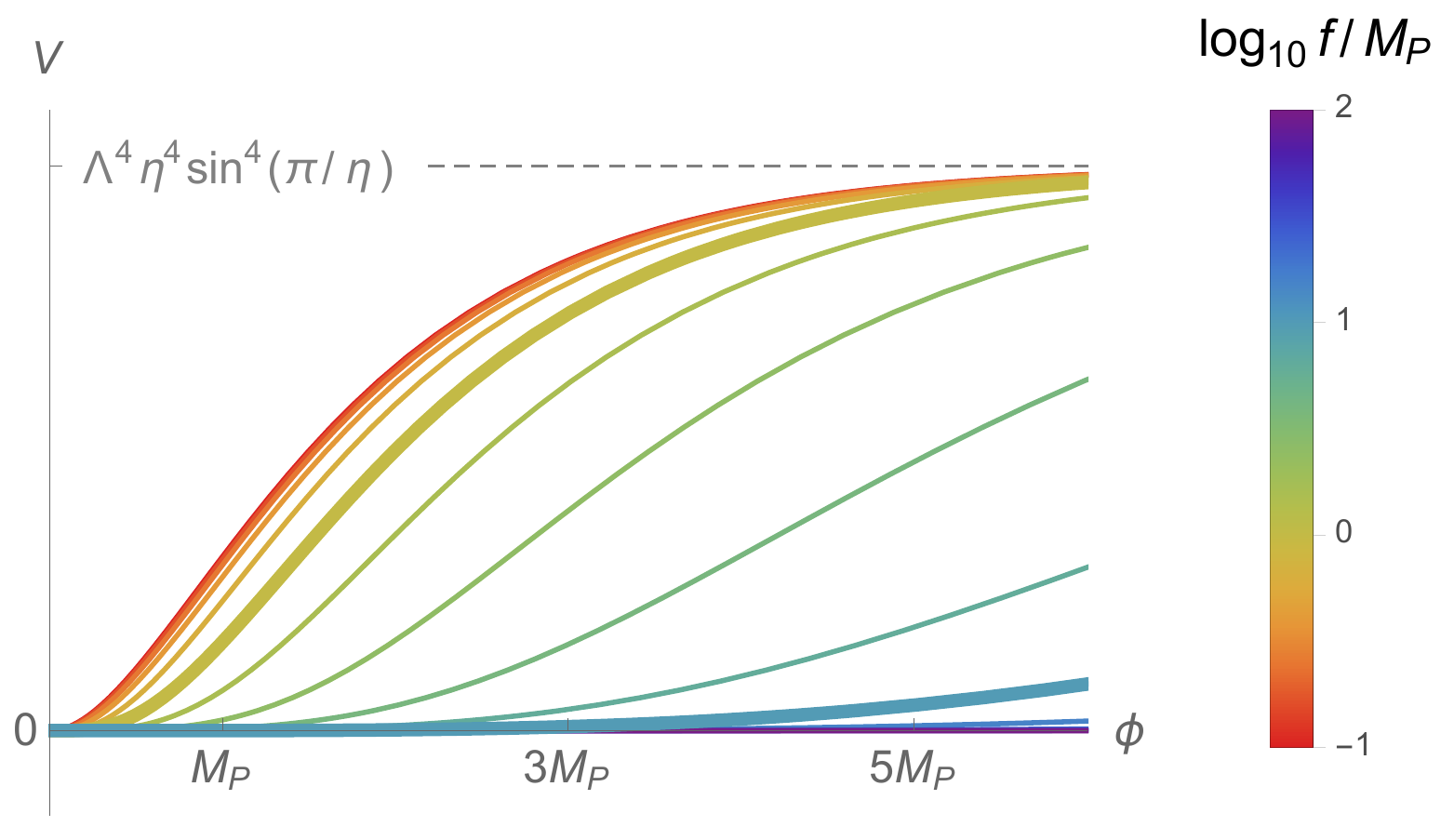}
\includegraphics[width=\columnwidth]{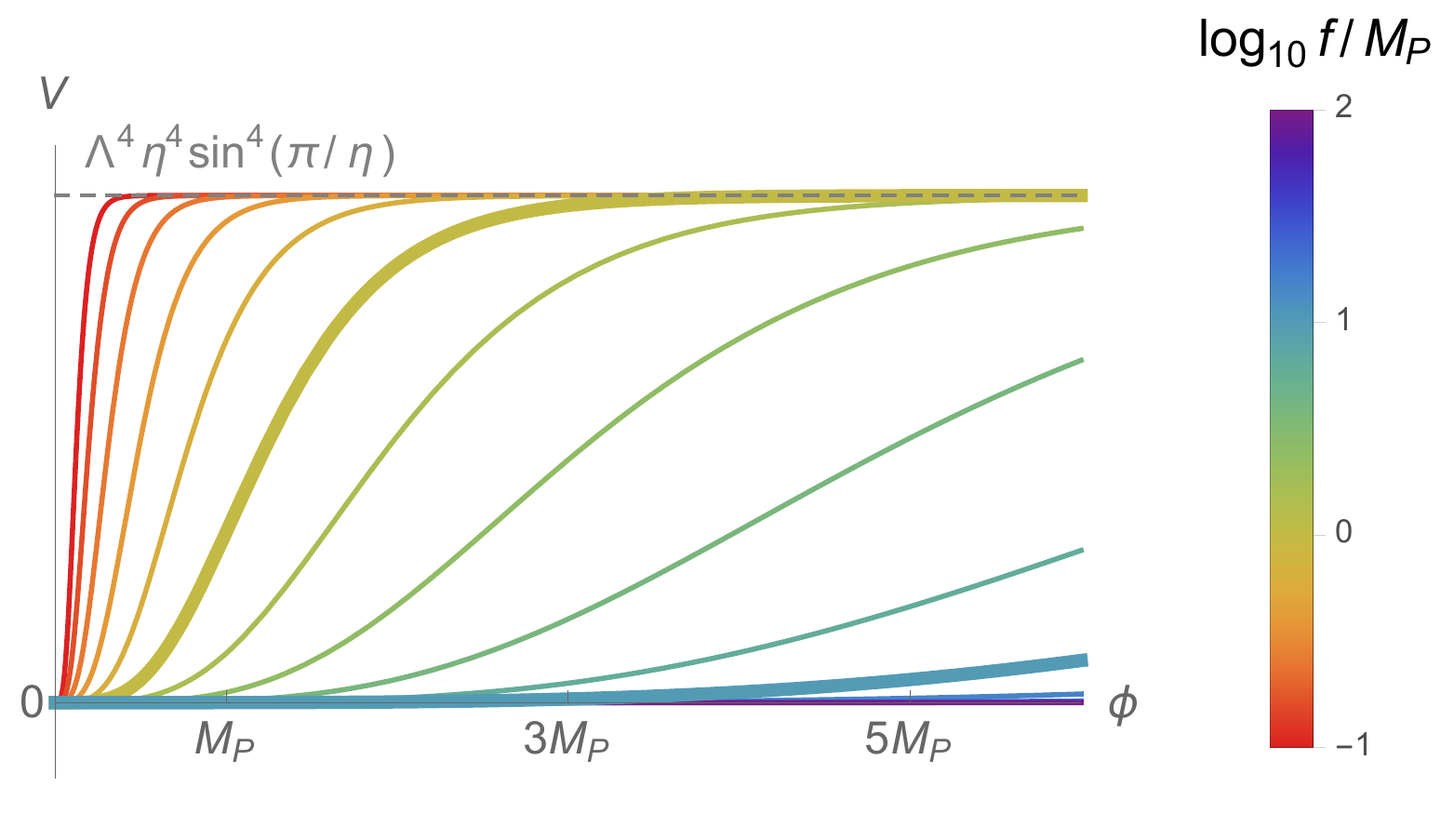}
\caption{\small
Einstein-frame potential as a function of the canonical inflaton $\phi$.
Note that $\phi = 0$ corresponds to $\phi_\J = \pi f$,
while $\phi = \infty$ corresponds to $\phi_\J = + 0$
(see Fig.~\ref{fig:SetupJordan} and Eq.~(\ref{eq:dphidphiJMetric}) or (\ref{eq:dphidphiJPalatini})).
(Left) The metric formulation with the kinetic term $\kappa = 1$ and $\eta = 2$.
(Right) The Palatini formulation with $\eta = 2$.
In both panels, the upper and lower thick lines correspond to $f=\MP$ and $f=10\MP$, respectively.
}
\label{fig:SetupEinstein2}
\end{center}
\end{figure*}

For the exponentially flat potential~(\ref{eq:VapproxMetric_kappa=0}), 
the slow-roll parameters at the leading order in $N$ become
\begin{align}
\epsilon_V
&\simeq
\frac{3}{4N^2},
~~~~
\eta_V
\simeq 
- \frac{1}{N},
\end{align}
and therefore the inflationary predictions are given by
\begin{align}
n_s
&\simeq
1 - \frac{2}{N},
~~~~
r
\simeq 
\frac{12}{N^2}.
\label{eq:nsrMetric_kappa=0}
\end{align}

\subsubsection*{Case $\kappa = 1$}

We next consider the case with the inflaton kinetic term ($\kappa = 1$).
In this case we do not show the analytic solution to Eq.~(\ref{eq:dphidphiJMetric}) since it is rather lengthy.
The kinetic term in the Einstein frame has contributions both from the original and new ones,
and as a result the Einstein-frame potential as a function of the canonical inflaton depends on the value of $f$.
However, as long as the new kinetic term $\sim (d\ln\Omega/d\phi_\J)^2$ dominates in Eq.~(\ref{eq:dphidphiJMetric}),
we expect that the potential is exponentially stretched in the horizontal direction
and the resulting predictions become the attractor values.

The top panel of Fig.~\ref{fig:SetupEinstein2} is the Einstein-frame potential with $\eta = 2$ for various choices of $f$.
The two thick lines in the figure correspond to $f = \MP$ (upper) and $f = 10\MP$ (lower).
We see that the potential shape is almost quadratic for larger $f$ $(\gtrsim 1)$, 
while it becomes exponentially flat for smaller $f$ $(\lesssim 1)$.
Note that the asymptotic shape $f \to 0$ is identical to $\eta = 2$ in Fig.~\ref{fig:SetupEinstein1},
since in this limit the original kinetic term becomes negligible.

\subsection*{Palatini formulation}

In the Palatini formulation, the Ricci tensor $R_{\J  \mu \nu}$ is no more a function of the metric $g_{\J  \mu \nu}$
but instead regarded as a function of the connection (symbolically denoted as $\Gamma$).\footnote{
In this paper we assume that the connection is torsion-free for simplicity.
}
As a result, the metric redefinition (\ref{eq:metric}) works only on the first factor of 
$R_\J = g_\J^{\mu \nu} R_{\J  \mu \nu} (\Gamma)$:
\begin{align}
R_\J
&=
\Omega
R.
\end{align}
Note that this redefinition does not give rise to a new kinetic term.
The relation between the original inflaton $\phi_\J$ and new inflaton $\phi$ is now given by
\begin{align}
\frac{d\phi}{d\phi_\J}
&=
- \sqrt{\frac{1}{\Omega}},
\label{eq:dphidphiJPalatini}
\end{align}
with the boundary condition $\phi = 0$ at $\phi_\J = \pi f$.
Here we put the minus sign because of the same reason as the metric case.
We can explicitly solve Eq.~(\ref{eq:dphidphiJPalatini}) and obtain
\begin{align}
\phi
&=
- \eta f \sin \left( \frac{\pi}{\eta} \right)
\log
\left[
\frac{\tan(\phi_\J / 2\eta f)}{\tan(\pi / 2\eta)}
\right].
\end{align}
Note that we identified $\phi = 0$ with $\phi_\J = \pi f$ to obtain this expression.
Since $\Omega$ is quadratic in $\phi_\J$ around $\phi_\J \sim +0$, 
the relation between the two fields (\ref{eq:dphidphiJPalatini}) gives 
$\alpha \phi \sim -\ln \phi_\J$ or $\phi_\J \sim e^{- \alpha \phi}$ with $\alpha = \sqrt{\omega}/\eta f = 1/\eta f \sin (\pi/\eta)$.
On the other hand, the Einstein-frame potential approaches to a constant value $V \to \Lambda^4 \eta^4/\omega^2$ 
for $\phi_\J \to +0$ as shown in Fig.~\ref{fig:SetupJordan}.
Therefore, in terms of the canonical field $\phi$, 
the Einstein-frame potential is exponentially ``stretched" in the horizontal direction around $\phi_\J \sim +0$.

In fact, the exact potential shape becomes
\begin{align}
V
&= 
\Lambda^4
\cos^8 \left( \frac{\pi}{2\eta} \right)
\nonumber \\
&~~~~
\times
\left[
e^{\frac{1}{\eta \sin(\pi/\eta)} \frac{\phi}{f}}
+
\tan^2 \left( \frac{\pi}{2\eta} \right)
e^{- \frac{1}{\eta \sin(\pi/\eta)} \frac{\phi}{f}}
\right]^4
\nonumber \\
&~~~~
\times
\sin^4
\left[
2\eta
\arctan 
\left[
\tan \left( \frac{\pi}{2\eta} \right)
e^{- \frac{1}{\eta \sin(\pi/\eta)} \frac{\phi}{f}}
\right]
\right].
\label{eq:VPalatini}
\end{align}
The potential for $\eta = 2$ is plotted in the right panel of Fig.~\ref{fig:SetupEinstein2} for various values of $f$.
It is again instructive to expand the potential assuming $e^{- \phi / \eta \sin(\pi / \eta) f} \ll 1$:
\begin{align}
V
&\simeq
\Lambda^4 \eta^4
\sin^4 \left( \frac{\pi}{\eta} \right)
\nonumber \\
&~~~~
\times
\left[
1
-
\frac{8}{3} (\eta^2 - 1)
\tan^2 \left( \frac{\pi}{2\eta} \right)
e^{- \frac{2}{\eta \sin(\pi/\eta)} \frac{\phi}{f}}
+
\cdots
\right].
\label{eq:VapproxPalatini}
\end{align}

\begin{figure*}
\begin{center}
\includegraphics[width=\columnwidth]{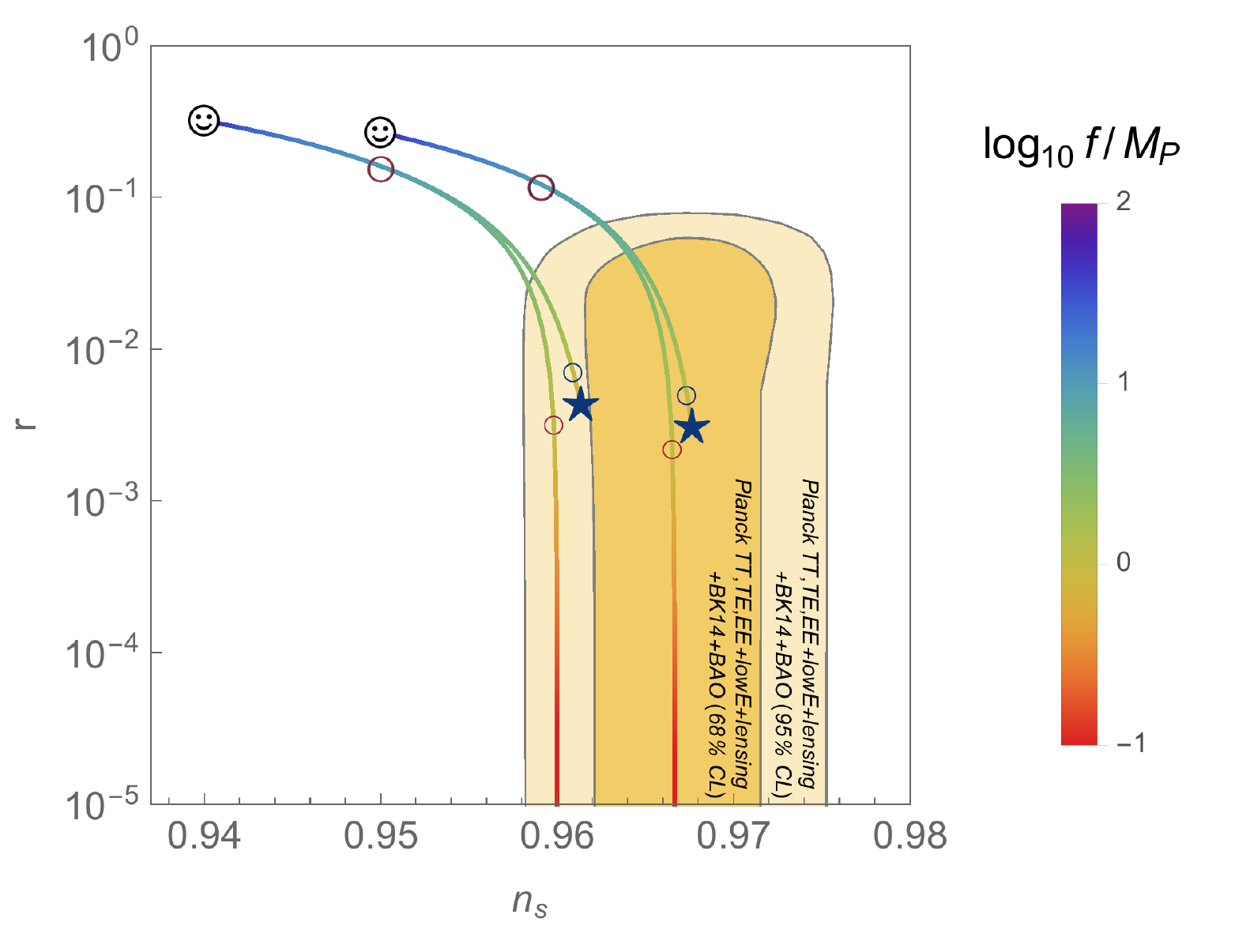}
\includegraphics[width=\columnwidth]{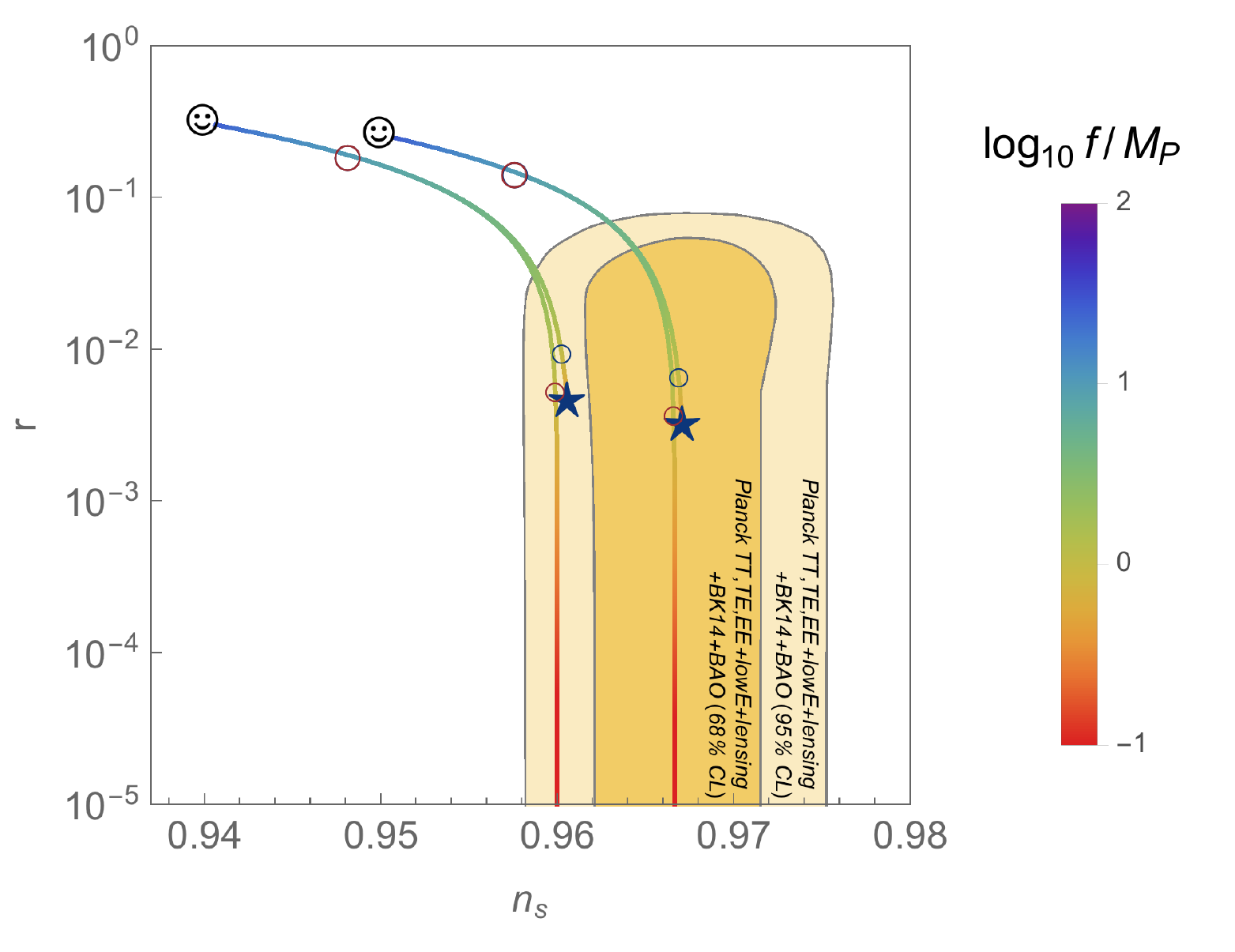}
\caption{\small
Inflationary predictions of the setup (\ref{eq:Setup})--(\ref{eq:Omega})
for $\eta = 2$ (left) and $3$ (right).
The left lines are for $N = 50$ while the right ones are for $N = 60$.
The metric formulation with the kinetic term in the Jordan frame ($\kappa = 1$) predicts the lines 
starting from the smiley markers and ending at the stars.
In the absence of the kinetic term ($\kappa = 0$),
the prediction comes at the position of the stars independently of $f$.
The Palatini formulation predicts the lines going into the $r \ll 1$ region.
The value of $f$ is also shown as a color plot, 
while $f = \MP$ and $f = 10\MP$ are indicated by the small and large open circles, respectively.
The blue (red) open circles correspond to the metric (Palatini) formulations.
The large open circles for both formulations almost overlap with each other.
}
\label{fig:nsr_eta=2&3}
\end{center}
\end{figure*}

We see that the potential develops a plateau for $\phi \gg 1/\alpha = \eta f \sin(\pi / \eta)$.
Note that, in contrast to the metric formulation, the potential becomes steeper and steeper as $f$ decreases.
This can be seen by comparing the exponent of Eqs.~(\ref{eq:VapproxMetric_kappa=0}) and (\ref{eq:VapproxPalatini}).
Also, for $\eta \to \infty$, the potential (\ref{eq:VPalatini}) becomes simpler
\begin{align}
V
&\to
\Lambda^4 
e^{\frac{4\phi}{\pi f}}
\sin^4
\left[
\pi
e^{- \frac{\phi}{\pi f}}
\right]
~~
{\rm for}
~~
\eta \to \infty.
\label{eq:VetainfPalatini}
\end{align}
We see that the potential (\ref{eq:VetainfPalatini}) indeed coincides with the potential (\ref{eq:VetainfMetric_kappa=0})
in the metric formulation with the choice $f = \sqrt{6}/\pi$.

For the exponentially flat potential~(\ref{eq:VapproxPalatini}), 
the slow-roll parameters at the leading order in $N$ become
\begin{align}
\epsilon
&\simeq
\frac{\eta^2 \sin^2 (\pi / \eta)}{8N^2} f^2,
~~~~
\eta
\simeq 
- \frac{1}{N},
\end{align}
and therefore the inflationary predictions are given by
\begin{align}
n_s
&\simeq
1 - \frac{2}{N},
~~~~
r
\simeq 
\frac{2 \eta^2 \sin^2 (\pi / \eta)}{N^2} f^2.
\label{eq:nsrPalatini}
\end{align}
The results coincide with those for the metric formulation with $\kappa=0$ in Eq.~\eqref{eq:nsrMetric_kappa=0}
when $f = \sqrt{6}/\eta \sin(\pi/\eta) \simeq \sqrt{6}/\pi$, which holds for $\eta \gg 1$.

\section{Observational predictions}
\label{sec:Results}

\begin{figure}[t]
\begin{center}
\includegraphics[width=\columnwidth]{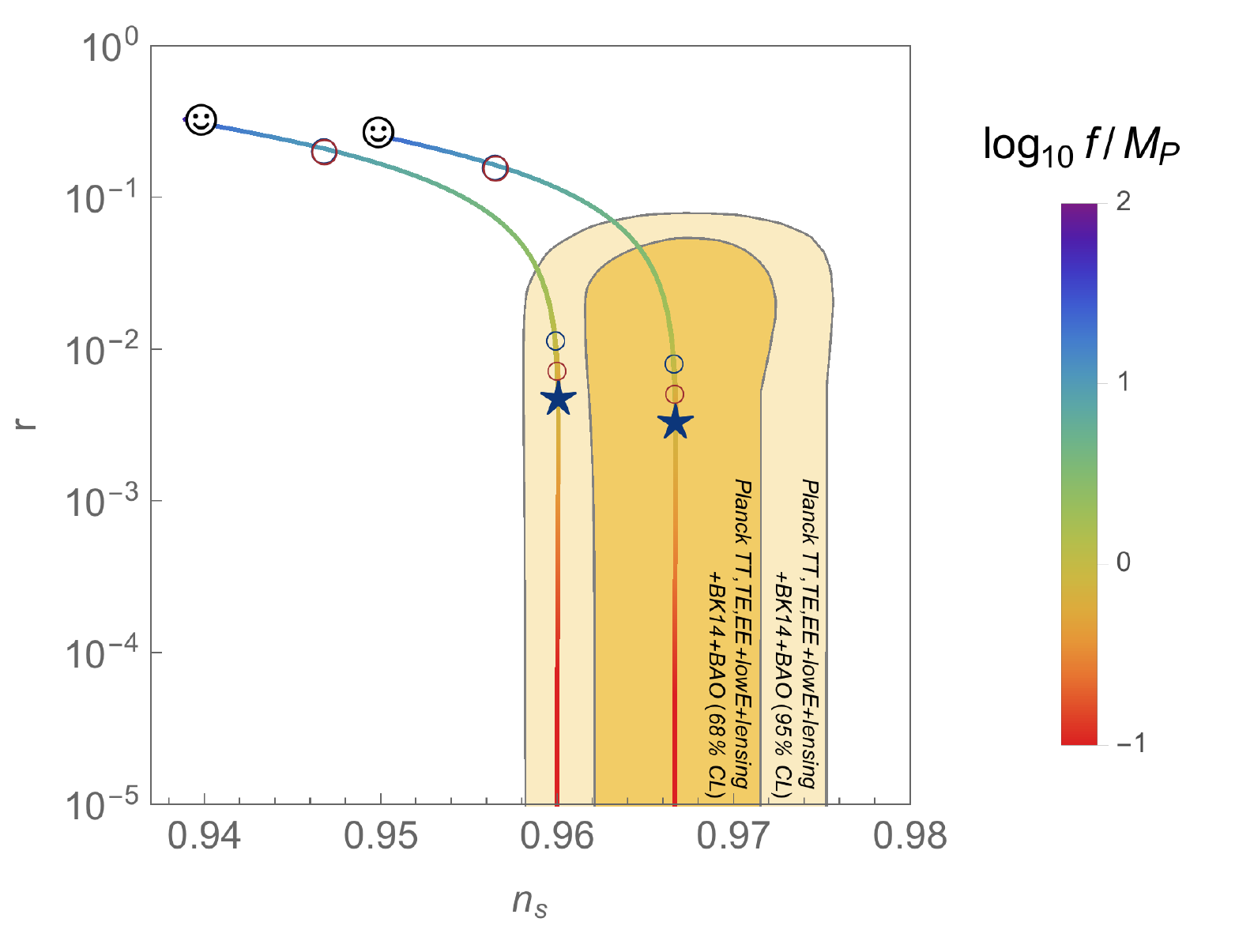}
\caption{\small
The same as Fig.~\ref{fig:nsr_eta=2&3}, except that $\eta = 10$. 
The small open circles are for $f = \MP$, while the large ones are for $f = 10\MP$. 
For the small open circles, the blue and red ones correspond to the metric and Palatini formulations, respectively,
while the large ones overlap for both formulations.
}
\label{fig:nsr_eta=10}
\end{center}
\end{figure}

In this section we present numerical results for the inflationary predictions in the setup explained in the previous section,
and also discuss possible reheating processes.

Figures~\ref{fig:nsr_eta=2&3} and \ref{fig:nsr_eta=10} are the inflationary predictions of the setup (\ref{eq:Setup}).
The former is for $\eta = 2$ (left panel) and $3$ (right panel), while the latter is for $\eta = 10$.
The lines starting from the smiley markers and ending at the blue stars 
(from $f = 10^2 \MP$ to $f = 10^{-1} \MP$) are 
for the metric formulation with the kinetic term ($\kappa = 1$).
The left and right lines correspond to $N = 50$ and $60$, respectively.
The value of $f$ is also indicated in color.

We see that, for the metric formulation,
the prediction is the same as that of the quartic chaotic inflation (smiley markers)
for larger $f$ ($\gtrsim 1$),
while it approaches to the attractor points (stars) for smaller $f$ ($\lesssim 1$).
For the metric formulation without the kinetic term ($\kappa = 0$),
the prediction is $f$-independent and always comes at the position of the stars.
We also note that their positions are almost $\eta$-independent,
as seen from the leading-order predictions (\ref{eq:nsrMetric_kappa=0}).

On the other hand, the lines going into the $r \ll 1$ region are for the Palatini formulation.
The prediction is again the same as that of the quartic chaotic inflation for larger $f$ ($\gtrsim 1$),
while it behaves differently from the metric case for smaller $f$ ($\lesssim 1$).
This behavior can be understood from Eq.~(\ref{eq:nsrPalatini}),
and therefore we see that, if $r$ is observationally found to much smaller than $10^{-2}$, 
the Palatini formulation can provide a better fit to the data.
We also see that the prediction with $f = \sqrt{6}/\pi$ in the Palatini formulation coincides with 
that of the metric formulation for $\eta \gg 1$.

Before moving on to conclusions, we comment on possible reheating mechanisms.
First, preheating~\cite{Kofman:1994rk,Kofman:1997yn} is very likely to occur in the current setup.
However, as expected from earlier studies, the dynamics is highly sensitive to 
(1) the existence of the kinetic term in the original frame~\cite{Ema:2016dny},
(2) the existence of other degrees of freedom 
(e.g. whether the inflaton is real or complex~\cite{Bezrukov:2008ut,GarciaBellido:2008ab,Ema:2016dny}, 
or whether the scalaron degree of freedom exists or not~\cite{Ema:2017rqn,He:2018gyf,He:2018mgb}),
(3) the choice of formulations~\cite{Rubio:2019ypq},
and so on.
All of these need further investigation, but are beyond the scope of the present paper.
Second, reheating via the perturbative inflaton decay can be implemented in both metric and Palatini formulations. 
In the current setup the inflaton actually becomes massless at the potential minimum in both formulations, 
and thus perturbative inflaton decay does not occur.\footnote{
This can be seen as the inflaton mass $m_\phi$ is given by
\begin{align*}
m_\phi^2 
&= 
K\left(\frac{\partial K}{\partial\phi_\J}\frac{\partial V_\J}{\partial\phi_\J}
+ K\frac{\partial^2V_\J}{\partial\phi_\J^2}\right),
\end{align*}
with $K(\phi_\J)\equiv d\phi_\J/d\phi$, and $\partial V_\J/\partial\phi_\J=\partial^2V_\J/\partial\phi_\J^2=0$ at $\phi_\J=\pi f$. 
Note that this is valid except for $\eta=2$ in the $\kappa=0$ case.
}
A remedy may be to incorporate a coupling to the SM Higgs field $H$. 
For instance, if we modify the Jordan-frame potential to include
\begin{align}
V_\J 
&\supset
\lambda \Lambda^2 |H|^2 \sin^2(\phi_\J/f),
\end{align}
with a certain coupling $\lambda$, the inflaton acquires a mass $m_\phi \sim \lambda \Lambda \langle H \rangle / f$ 
once the inflation ends and the electroweak symmetry is broken, 
whereas during inflation the Higgs is stabilized at the origin by this coupling
and thus the inflaton dynamics would be kept intact.\footnote{
The inflaton mass thus induced may fall within the reach of low energy experiments
such as the ones searching for Axion-like particle or dark photons. Further details are reserved for future study.
} 
Once the inflaton acquires a heavier mass than the Higgs, the reheating may take place by the inflaton decay into Higgs bosons.
Even if the tree-level decay is kinematically forbidden, 
the inflaton can decay into a pair of lighter SM particles through the Higgs loop diagrams.

\section{Conclusion}
\label{sec:Conclusion}

In this paper we discussed a new realization of cosmic inflation
where both the inflaton potential and conformal factor are periodic functions of the inflaton field. 
In particular, we focused on the realization both in the metric and Palatini formulations of gravity, 
adopting a specific type of the potential and conformal factor, namely, 
sinusoidal functions as a variant possibility of natural inflation.
We showed that our setup gives inflationary predictions well consistent with cosmic microwave background observations in both formulations.
We also argued that, for the metric formulation case, the existence of the kinetic term in the Jordan frame is not mandatory, and the consistent predictions can be obtained.
Future observations, particularly those sensitive to $r = {\cal O}(10^{-3})$, 
may be able to distinguish between the metric and the Palatini formulations 
in our setup~\cite{Matsumura:2013aja,Inoue:2016jbg,Delabrouille:2017rct}.

\section*{Acknowledgments}

The work of RJ was supported by IBS under the project code, IBS-R018-D1.
The work of SP was supported in part by the National Research Foundation of Korea (NRF) grant 
funded by the Korean government (MSIP) (No. 2016R1A2B2016112) and (NRF-2018R1A4A1025334).
The work of KK was supported in part by the DOE grant DE--SC0011842 at the University of Minnesota.

\appendix

\section{Limiting values of $\eta$}
\label{app:Limit}

In this appendix we discuss $\eta \to 1$ limit in the metric formulation with $\kappa = 0$.

\begin{figure}[t]
\begin{center}
\includegraphics[width=\columnwidth]{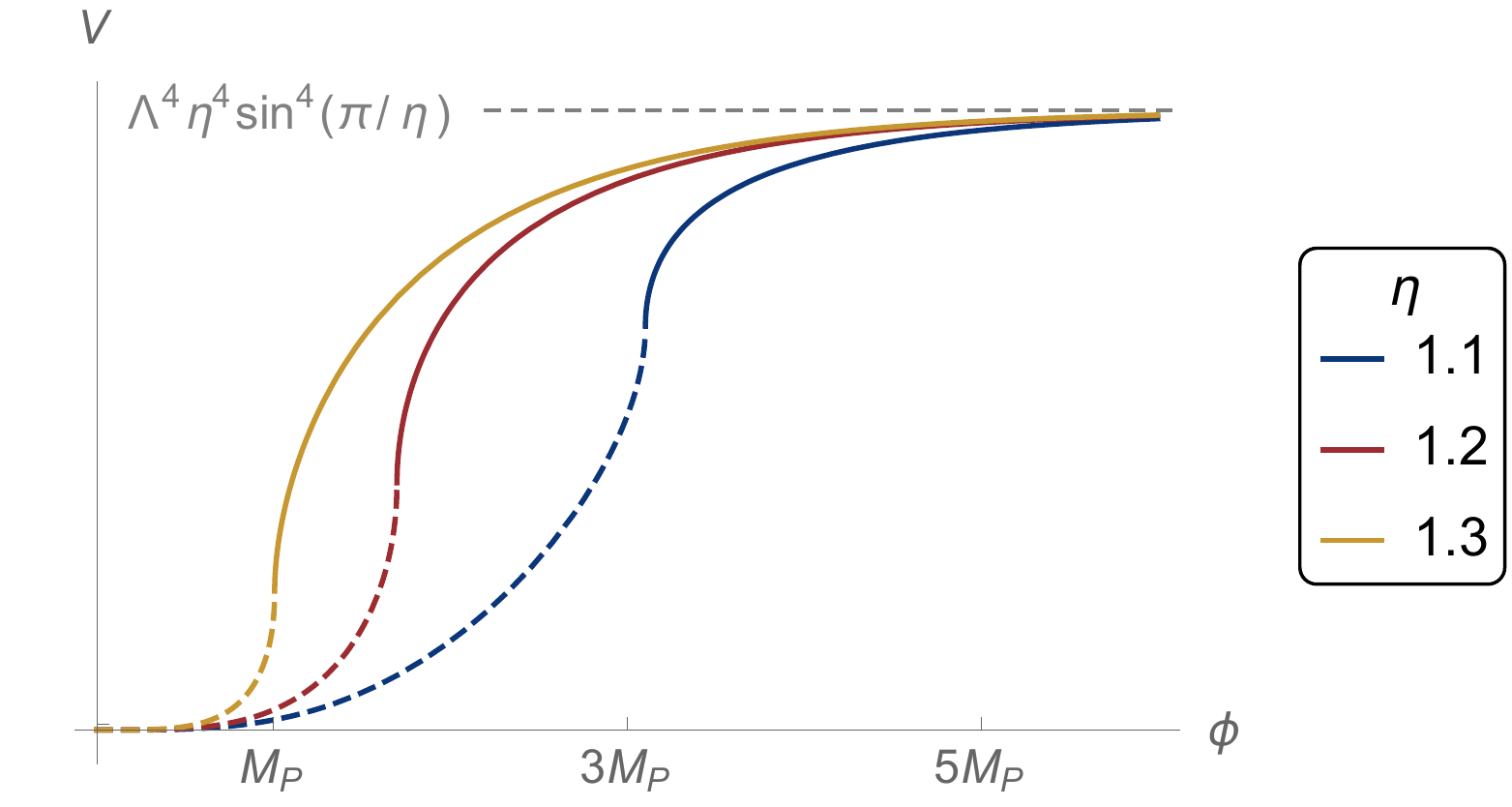}
\caption{\small
Inflaton potential $V$ for $\eta = 1.1$ (blue), $1.2$ (red) and $1.3$ (yellow) 
for the metric formulation without the kinetic term $\kappa = 0$.
}
\label{fig:Setup_Singular}
\end{center}
\end{figure}

\begin{figure}[t]
\begin{center}
\includegraphics[width=\columnwidth]{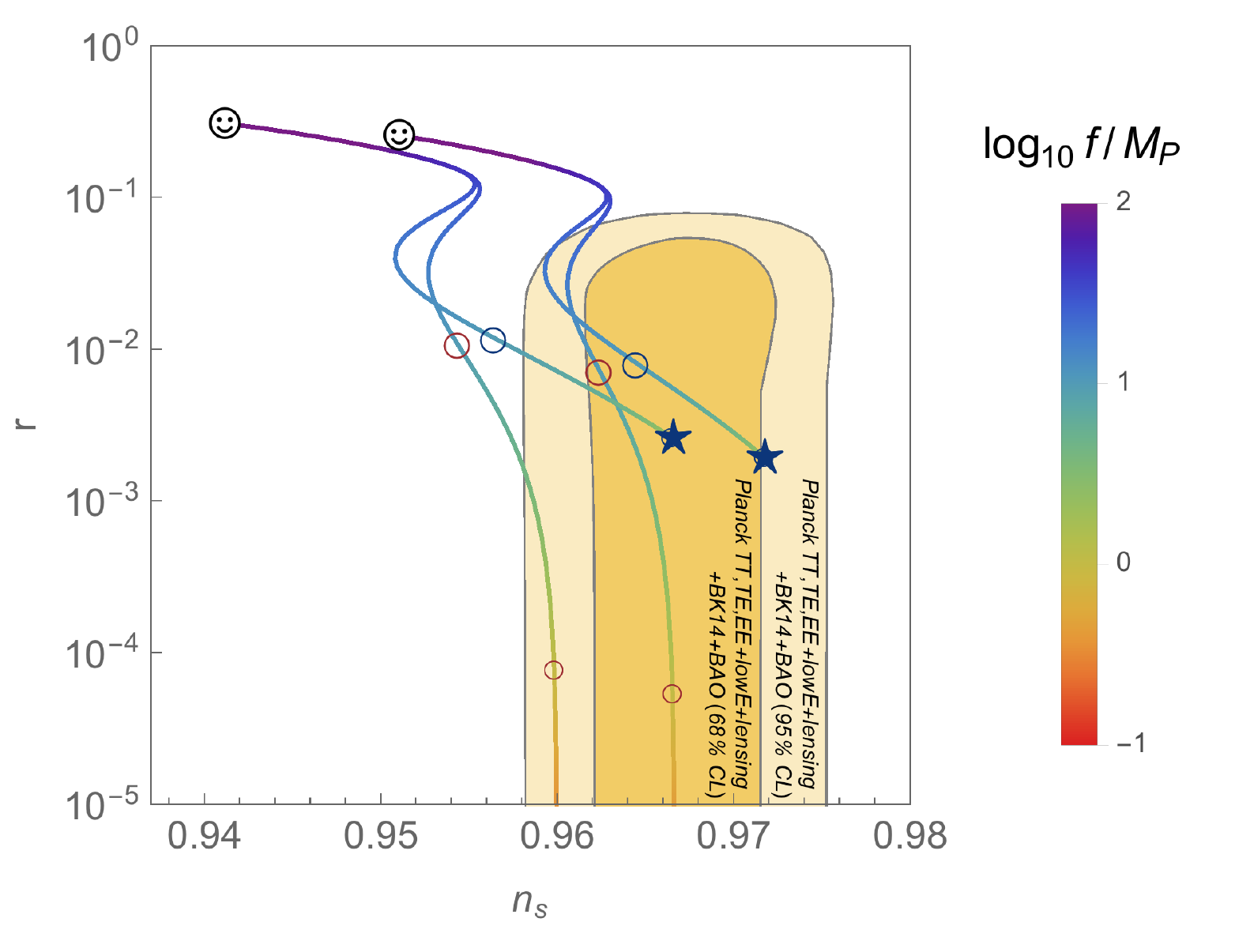}
\caption{\small
Inflationary predictions of the setup (\ref{eq:Setup})--(\ref{eq:Omega}) for $\eta = 1.1$. 
The small and large open circles indicate $f = \MP$ and $10\MP$, respectively.
}
\label{fig:nsr_Others}
\end{center}
\end{figure}

For $\eta < 2$, the conformal factor $\Omega$ takes its maximum 
between $\phi_\J = +0$ and $\phi_\J = \pi f$.
As a result, Eq.~(\ref{eq:dphidphiJMetric}) needs a special care for $\kappa = 0$.
We obtain
\begin{align}
V
&= 
\Lambda^4 
e^{2\sqrt{\frac{2}{3}} \phi}
\sin^8 \left( \frac{\pi}{\eta} \right)
\nonumber \\
&~~~~
\times
\sin^4
\left[
\eta
\arcsin
\left[
e^{-\frac{1}{2} \sqrt{\frac{2}{3}} \phi} \Big/ \sin \left( \frac{\pi}{\eta} \right)
\right]
\right],
\\
\Omega
&=
\frac{\sin^2 \left( \phi_\J / \eta f \right)}{\sin^2 \left( \pi/\eta \right)}
=
e^{- \sqrt{\frac{2}{3}} \phi},
\end{align}
for $0 < \phi_\J < \eta \pi f / 2$ (i.e. $\phi > - \sqrt{6} \log \left[ \sin \left( \pi / \eta \right) \right]$), 
while
\begin{align}
V
&= 
\Lambda^4 
e^{- 2\sqrt{\frac{2}{3}} \phi}
\sin^4
\left[
\eta\pi
-
\eta
\arcsin
\left[
\sin
\left(
\frac{\pi}{\eta}
\right)
e^{\frac{1}{2} \sqrt{\frac{2}{3}} \phi}
\right]
\right],
\\
\Omega
&=
\frac{\sin^2 \left( \phi_\J/f \right)}{\sin^2 \left( \pi/\eta \right)}
=
e^{\sqrt{\frac{2}{3}} \phi},
\end{align}
for $\eta \pi f / 2 < \phi_\J < \pi f$ (i.e. $0 < \phi < - \sqrt{6} \log \left[ \sin \left( \pi / \eta \right) \right]$).

Fig.~\ref{fig:Setup_Singular} is the inflaton potential $V$ 
for $\eta = 1.1$ (blue), $1.2$ (red) and $1.3$ (yellow) for the metric formulation 
without the kinetic term $\kappa = 0$.
We see that the potential is monotonic and continuous at the boundary value
$\phi_\J = - \sqrt{6} \log \left[ \sin \left( \pi / \eta \right) \right]$,
which corresponds to the junction of the solid lines to the dashed lines.
However, we also see that the derivative diverges at this point.
As long as we do not care about this divergence, 
we can calculate the inflationary predictions in the same procedure as the main text.
Fig.~\ref{fig:nsr_Others} is the inflationary predictions for $\eta = 1.1$.

\bibliography{ref}

\end{document}